\documentclass[preprint2,usenatbib]{emulateapj}

\newcommand{\rms}{{\emph{rms}}}
\newcommand{\kms}{km\,s$^{-1}$}                                 
\newcommand{\msec}{m\,s$^{-1}$}                                 
\newcommand{\sci}[2]{{:1}\!\times10^{:2}}                         

\newcommand{\ie}{i.e.}
\newcommand{\eg}{e.g.}

\newcommand{\cf}{{cf.}}

\newcommand{\topp}{\emph{top}}

\newcommand{\bott}{\emph{bottom}}

\newcommand{\rii}{\emph{right}}
\newcommand{\lee}{\emph{left}}
\newcommand{\panel}{\emph{panel}}
\newcommand{\panels}{\emph{panels}}
\newcommand{\inset}{\emph{inset}}
\newcommand{\insets}{\emph{insets}}
\newcommand{\smeisymb}{$\times$}
\newcommand{\wiresymb}{small dots}
\newcommand{\specsymb}{$\bullet$}
\newcommand{\cday}{c\,day$^{-1}$}
\newcommand{\mhz}{$\mu$Hz}

\newcommand{\aumi}{$\alpha$~UMi}
\newcommand{\aumii}{$\alpha$~Ursae~Minoris}  
\newcommand{\polaris}{Polaris}

\newcommand{\dinshaw}{D89}
\newcommand{\hatzes}{H00}
\newcommand{\spec}{{\sc ast}}
\newcommand{\virgo}{{\sc virgo}}
\newcommand{\wire}{{\sc wire}}

\newcommand{\smei}{{\sc smei}}

\newcommand{\nasa}{{\sc nasa}}

\newcommand{\coriolis}{{\sc coriolis}}
\newcommand{\hipp}{{\sc hipparcos}}

\newcommand{\Fei}{Fe\,{\sc i}}
\newcommand{\kmps}{km~s$^{-1}$}

\slugcomment{Accepted by ApJ - 2008 April 21}
\shorttitle{Polaris the Cepheid returns}
\shortauthors{Bruntt et al.}

\begin{document}


\title{Polaris the Cepheid returns:\\$4.5$ years of monitoring from ground and space}

\author{H.~Bruntt\altaffilmark{1} and N.~R.~Evans\altaffilmark{2} and D.~Stello\altaffilmark{1} and A.~J.~Penny\altaffilmark{3} and J.~A.~Eaton\altaffilmark{4} and D.~L.~Buzasi\altaffilmark{5} and D.~D.~Sasselov\altaffilmark{6} and H.~L.~Preston\altaffilmark{5} and E.~Miller-Ricci\altaffilmark{6}}

\altaffiltext{1}{School of Physics, University of Sydney, NSW 2006, Australia}
\altaffiltext{2}{Smithsonian Astrophysical Observatory, 60 Garden St., Cambridge, MA 02138, USA}
\altaffiltext{3}{University of St Andrews, School of Physics and Astronomy, North Haugh, St Andrews, KY16 9SS, United Kingdom}
\altaffiltext{4}{Center of Excellence in Information Systems, Tennessee State University, Nashville, TN 37203, USA}
\altaffiltext{5}{Department of Physics, US Air Force Academy, Colorado Springs, CO 80840, USA}
\altaffiltext{6}{Astronomy Department, Harvard University, 60 Garden St., Cambridge, MA 02138, USA}

\begin{abstract}
We present the analysis of $4.5$ years of nearly continuous 
observations of the classical Cepheid \polaris, 
which comprise the most precise data available for this star.
We have made spectroscopic measurements from ground and photometric measurements from 
the \wire\ star tracker and the \smei\ instrument on the \coriolis\ satellite. Measurements 
of the amplitude of the dominant oscillation ($P=4$\,d), that go back more than a century, 
show a {\it decrease} from $A_V=120$ mmag to $30$ mmag around the turn of the millennium.
It has been speculated that the reason for the decrease in
amplitude is the evolution of \polaris\ towards the edge of the instability strip.
However, our new data reveal an {\it increase} in the amplitude by $\sim30\%$ from 2003--2006.
It now appears that the amplitude change is cyclic rather than monotonic,
and most likely the result of a pulsation phenomenon.
In addition, previous radial velocity campaigns have claimed the detection 
of long-period variation in \polaris\ ($P>40$\,d). 
Our radial velocity data are more precise than previous datasets,
and we find no evidence for additional variation for periods 
in the range 3--50\,d with an upper limit of 100\,\msec.
However, in the \wire\ data we find evidence of variation 
on time-scales of 2--6 days, which we interpret as being due to granulation.
\end{abstract}





\section{Introduction\label{sec:intro}}

In addition to being arguably the most famous and in practice useful star
other than the Sun, \polaris\ has a number of properties 
that may provide insights that are important to stellar astrophysics.  
It is the nearest and brightest classical Cepheid, 
oscillating in a single mode of pulsation with a period around four days.
The \hipp\ parallax constrains its luminosity and 
allowed \cite{feast97} to argue that the mode of pulsation 
must be the first overtone, which is upheld by the recent
reevaluation of the \hipp\ data \citep{leeuwencepheid07}.
As discussed by \cite{evans02}, \polaris\ has a number of 
unusual pulsation properties, including a very 
small pulsation amplitude, and, as for other overtone pulsators, 
it has a rapid period change.  It has been clear from studies as far
back as \cite{par56} and \cite{sza77} that overtone 
pulsators (``s Cepheids" as they were then called) had 
rapid period changes,  more rapid than can be 
explained by evolution during a second or third crossing of the instability strip.  

It was found by \cite{ferro83} that the main period 
of \polaris\ is increasing (316\,s per century) and that its peak-to-peak 
amplitude has decreased significantly, from about 140 to 70 mmag (Johnson $B$ filter), 
based on photoelectric measurements collected in the period 1930--1980.
\cite{kamper98} analysed radial velocity measurements from 1900 to 1998
and could confirm the decrease in amplitude.
However, their radial velocity data from the end of the period 
showed that the decrease had apparently stopped.
\cite{turner05} analysed both radial velocity and photometry data and 
found evidence for a sudden change in oscillation period around 1963--1966,  
and that the amplitude change became steeper at the same time.

Our aim is to shed light on the properties of the oscillation of \polaris,
based on the analysis of a nearly continuous dataset spanning $4.5$ years.
We have photometry from two satellite missions and simultaneous 
spectroscopic monitoring from the ground, each dataset
being superior to previous data available for \polaris.

%
\begin{table}
\caption{Observing log. The point-to-point
scatter, $\sigma$, is given in mmag (\wire\ $+$ \smei\ photometry) and \kms\ (spectroscopy).}
\label{tab:data} \centering
\begin{tabular}{l l l r r | l} \hline \hline
       &            &                 & \multicolumn{1}{c}{}                    & \multicolumn{1}{c}{Data}    &          \\
Source & Date start & Date end        & \multicolumn{1}{c}{ $T_{\rm obs}$\,[d]}  &  \multicolumn{1}{c}{points} & $\sigma$ \\ \hline
\wire  & 2004 Jan 17  & 2004 Feb 15   &   29.0 &  41\,162    & 0.88  \\    
\wire  & 2004 Jul 10  & 2004 Jul 30   &   20.2 &  33\,937    & 0.30  \\    
\wire  & 2005 Jan 31  & 2005 Feb 12   &   12.6 &  14\,240    & 0.29  \\    
\smei  & 2003 Apr 6   & 2006 Dec 31   & 1365.4 &  13\,543    & 6.4   \\  \hline  
\spec  & 2003 Dec 25  & 2007 Oct 26   & 1401.0 &      517    & 0.15  \\    
\dinshaw&1987 May 6   & 1988 Jan  3   &  241.0 &      175    & 0.44  \\  
\hatzes &1991 Nov 21  & 1993 Aug  2   &  640.4 &       42    & 0.07  \\  

\hline 

\end{tabular}
\end{table}
%

   \begin{figure*}
   \centering
   \includegraphics[width=14.5cm]{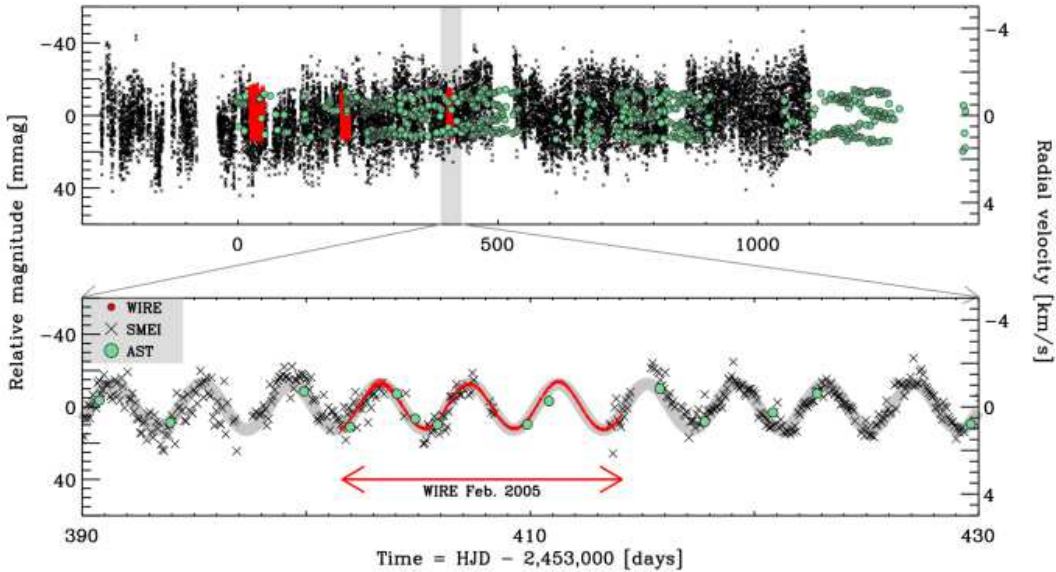}
      \caption{
 The photometry datasets from \smei\ and \wire\ and the radial velocities from \spec. 
 The \bott\ \panel\ is an expanded view of 40 days of observations 
 surrounding the last run with \wire.
 The thick gray curve is the fit of a single sinusoid to the \smei\ data.
         \label{fig:lc}}
   \end{figure*}
%

\section{Observations and data reduction}
\label{sec:obs}

\polaris\ (\aumii) was observed with the star tracker on the Wide-field InfraRed Explorer (\wire; \citealt{bruntt07wire}) satellite
in January and February 2004, July 2004, and February 2005.
These runs lasted about 4, 3 and 2 weeks, respectively.  
In addition, \polaris\ was monitored using 3.8 years of
nearly continuous photometry from the \smei\ instrument on the \coriolis\ satellite.
These observations were obtained between April 2003 and the end of 2006.
We further used the 2\,m Tennessee State University Automatic Spectroscopic Telescope (AST; \citealt{eaton04, eaton07})
to collect 517 high dispersion spectra of \polaris\ over a period of $3.8$ years, 
from late 2003 to late 2007. 
A log of the photometric and radial velocity observations is given in Table~\ref{tab:data}.
We also list datasets from two previous radial velocity campaigns that
we have used for comparison in our analysis in Sect.~\ref{sec:rv} 
(D89: \citealt{dinshaw89} and H00: \citealt{hatzes00}).

The complete photometric light curve from \smei\ and \wire\ 
and the radial velocity data from \spec\ are shown 
in the \topp\ \panel\ in Fig.~\ref{fig:lc}.
Note that different units for the photometry (mmag) and velocities (\kms)
are given on the left and right abscissa, respectively.
The right abscissa is adjusted 
by the ratio of the measured amplitudes in the \spec\ spectroscopy 
and the \smei\ photometry. 
The \bott\ \panel\ in Fig.~\ref{fig:lc} shows the details
of the variation during 40 days.
It shows the last run done with \wire\ and illustrates 
the typical coverage with \smei\ and the \spec.
The thick gray curve is the fit of a single sinusoid to the complete \smei\ dataset.


%
\begin{table}
\caption{Times of maximum light in the \smei\ and \wire\ datasets.}
\label{tab:oc} \centering
\begin{tabular}{ccc}
\hline \hline
\multicolumn{1}{c}{$t_E$}               & \multicolumn{1}{c}{Epoch}    &  Source      \\ 
\multicolumn{1}{c}{[HJD\,$-2,453,000$]} &  &  \\ \hline


$ 391.327 $ & 6329  & \smei \\
$ 395.339 $ & 6330  & \smei \\
$ 399.387 $ & 6331  & \smei \\
$ 403.323 $ & 6332  & \wire \\
$ 407.274 $ & 6333  & \wire \\
$ 411.262 $ & 6334  & \wire \\
$ 415.418 $ & 6335  & \smei \\
$ 419.160 $ & 6336  & \smei \\
$ 423.215 $ & 6337  & \smei \\
$ 427.152 $ & 6338  & \smei \\

\hline 
\end{tabular}
\end{table}
%

The \wire\ dataset consists of about three million CCD stamps
extracted from the 512$\times$512 CCD SITe star tracker camera.
Each window is 8$\times$8 pixels and we carried out aperture photometry
using the pipeline described by \cite{bruntt05}. 
The resulting point-to-point scatter range from $0.3$ to $0.9$ mmag (see Table~\ref{tab:data}). 
The noise is somewhat higher than the Poisson noise, 
\eg\ the high noise in the first \wire\ dataset is due to a high background sky level.

Data from the \spec\ consist of \'echelle spectra covering 
the wavelength range $5\,000$--$7\,100$ \AA\ at a resolution of about $30\,000$.
The velocities are derived from 
the correlation between the observed spectrum and a list of 74 mostly \Fei\ lines.
More details about the pipeline used to extract the radial velocities are given by \cite{eaton07}.
We find that the velocities deviate slightly from
the IAU velocity system ($\Delta v=-0.35\pm0.09$\,\kmps).
The drift of the velocities due to the motion of \polaris\ in its binary orbit \citep{kamper96}
was subtracted before the time-series analysis was carried out.

The Solar Mass Ejection Imager (\smei) \citep{eyles03, jackson04} is a
set of three cameras mounted on the side of the \coriolis\ satellite.
Each camera covers a $3\times60$ degree strip of sky and this is
projected onto $1260\times40$ pixels of a CCD. As the nadir-pointing
satellite orbits every 101 minutes in its Sun-Synchronous polar orbit
the cameras take continuous 4-second exposures, which result in roughly
4,500 images per orbit covering most of the sky. After bias and dark removal and
flat-fielding processing, the pixels about the target star are selected
from about ten images that contain the star for each orbit. These pixels are
then aligned and a standard PSF is fitted by least-squares.
Thus, the result is one brightness measure every 101 minutes.
There are problems with particle hits and a complex PSF, which shows
temporal variations that are not fully understood.
The short-term ($t<1.0$\,d) accuracies
are of the order of a few mmag, but there are longer term systematic
errors of the order of 10 mmag on the timescales of days and months,
which are not yet well understood. Allowing for time allocated for
calibrations and satellite operation, a time coverage of about 85\% was
maintained.

   \begin{figure}
   \centering
   \includegraphics[width=8.8cm]{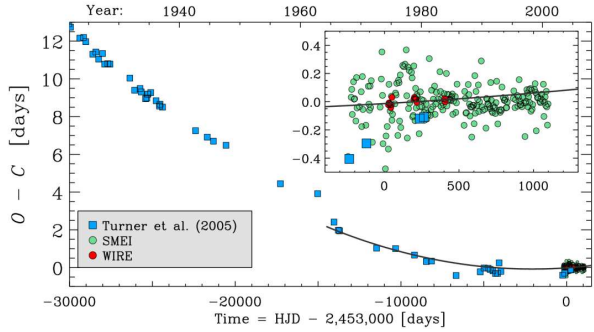}
      \caption{$O-C$ diagram. The solid line is a fit to the data taken after 1965. 
               The \inset\ shows the details of the new $O-C$ data from \smei\ and \wire.
         \label{fig:oc}}
   \end{figure}
%

%
\section{Period change of the main mode\label{sec:oc}}
%

Observed epochs of maximum light in \polaris, spanning more than a century, 
have indicated a significant change in period \citep{ferro83, fernie93, turner05}. 
These studies have considered $O-C$ diagrams\footnote{$O-C$ diagrams 
display observed times of maximum light minus calculated times, 
assuming a constant period, plotted vs.\ time.} 
and fitted a parabola, which is equivalent 
to assuming the period changes linearly with time. 
We have combined new measured epochs with those from \cite{turner05}.
A sudden change in the period has been noted in the 1960's and
therefore we have only used epochs observed since 1965.


We measured 223 and 16 epochs of maximum light in the \smei\ and \wire\ datasets, respectively.
From the $O-C$ analysis by \cite{ferro83} we expected the period to 
decrease by 14 seconds over the $4.5$ years time span of our dataset.
Thus, when predicting the epochs of maximum light we could safely assume a constant period.
The epochs were predicted by fitting a single sinusoid to the \smei\ data. 
The fit is of the form $S(t)=A \sin [2 \pi (f\,(t-t_0) + \phi) ]$, where
$A$ is the amplitude\footnote{Previous studies on \polaris\ have used peak-to-peak amplitudes.
We use amplitudes unless otherwise specified.}, $t$ is the time (Heliocentric Julian Date), 
$t_0$ is the zero point, $P=1/f$ is the period, and $\phi$ is the phase. 
The result is $P = 3.972111\pm0.000054$~days and 
phase $\phi=0.2158\pm0.0022$ for the zero point $t_0=2,530,000$.   
This fit is shown with a thick gray line in the \bott\ \panel\ in Fig.~\ref{fig:lc}.
The uncertainties were determined from simulations as discussed in Sect.~\ref{sec:uncert}.


At each epoch, $E$, we selected all data points within half a period ($|t-E*P|<P/2$).
For the \smei\ data we required that at least 30 data points 
were available with at least 10 data points before and after the maximum. 
To estimate the time of maximum light, we fitted these data with a sinusoid 
with fixed frequency and amplitude, while the phase was a free parameter.
In Table~\ref{tab:oc} we list these times, including the epoch number
following the definition by \cite{turner05}, and the source of the data. 
The complete table is available in the on-line version of the paper, while
the times listed here correspond to the time interval covered in the \bott\ \panel\ in Fig.~\ref{fig:lc}.

Since the quality of the three datasets are very different, 
we computed weights based on the uncertainty of the epoch times, $t_E$.
This was determined by calculating the point-to-point scatter
after subtracting a parabola fitted to the $O-C$ data of each dataset.
The uncertainty on $t_E$ from \cite{turner05} is $\sigma(t_E) = 0.18$\,d while 
for \smei\ and \wire\ data the uncertainties are 0.12\,d and 0.014\,d.
Finally, we calculate the $O-C$ values using weights, $1 / \sigma^2(t_E)$,
and the most accurate value of the period, which is the one from \smei:
$O-C = (t_E-t_0) - (E-E_0) \cdot P_{\rm SMEI}$.
We chose the reference epoch to be $E_0=6333$, and using the fit to the \smei\ data 
found above, this corresponds to the time $t_0 = 2,453,407.277\pm0.009$.

Our final $O-C$ diagram is shown in Fig.~\ref{fig:oc}.
The recent data continue to match the smooth period change as indicated by the weighted parabolic fit,
corresponding to a period increase of $dP/dt=365\pm27$~sec/century. 
We note that the most recent \smei\ data lie systematically below the fit, which is
tightly constrained by the extremely well-defined times from \wire.
Note that if we only use data more recent than 1985, 
we conclude that the period has not changed, \ie\ $dP/dt = -19\pm95$~sec/century.


%
   \begin{figure*}
   \centering
   \includegraphics[width=14cm]{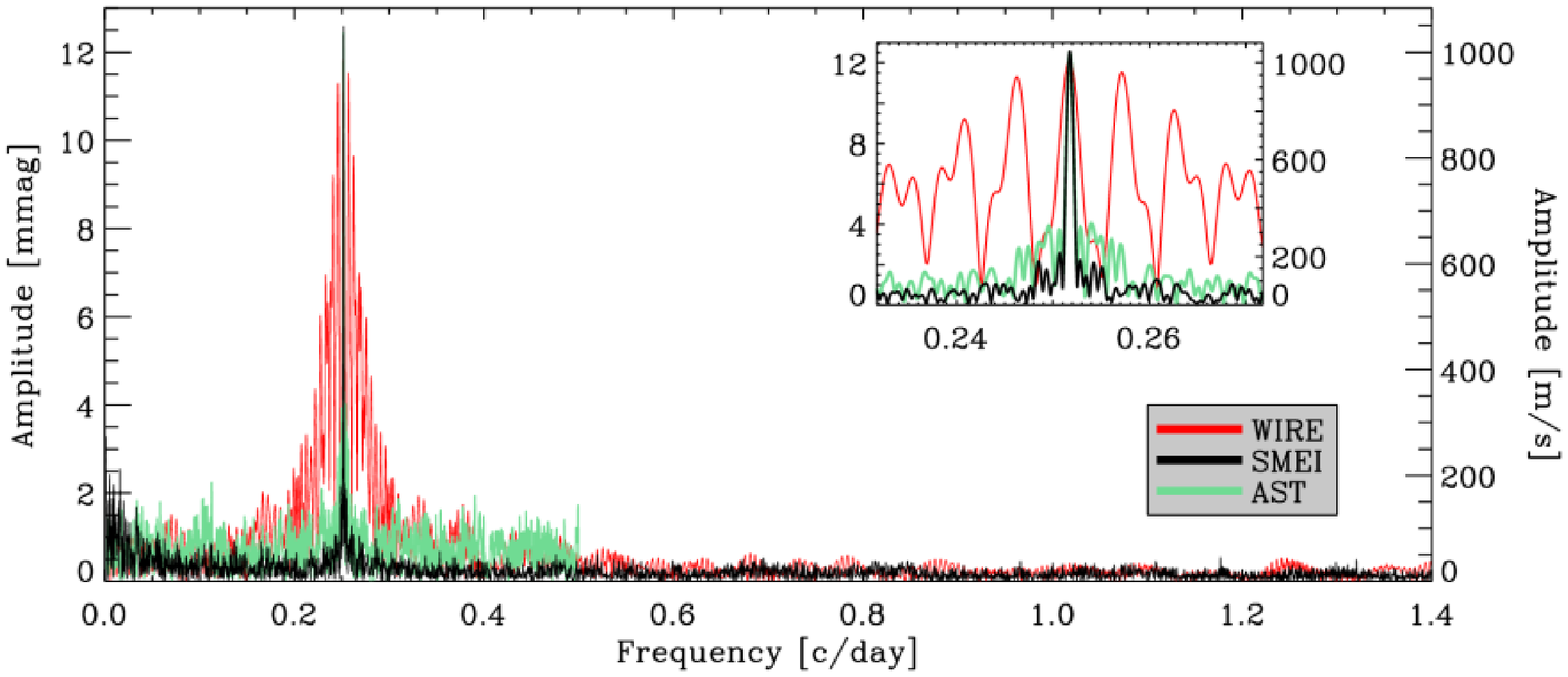}
   \includegraphics[width=14cm]{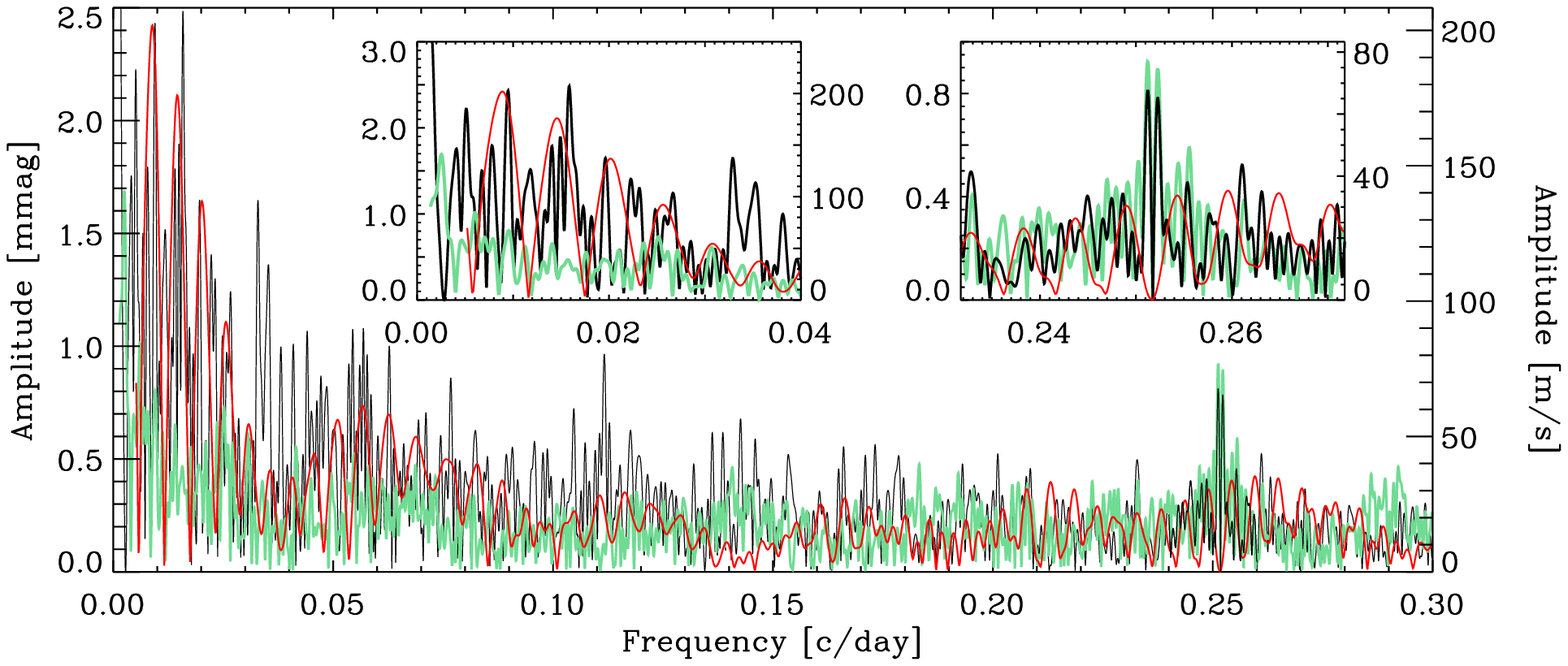}
      \caption{Amplitude spectra of the \spec\ radial velocity data 
and the photometric data from \smei\ and \wire. 
The \topp\ \panel\ is for the raw data and the \bott\ \panel\ is after
subtracting the main mode at $\simeq0.25$\,\cday.
The \insets\ show details of the amplitude spectra.
         \label{fig:amp}}
   \end{figure*}
%

\begin{table}
\caption{Results of the frequency analysis for the five datasets.
\label{tab:per}} \centering
\begin{tabular}{llcc} \hline \hline
Source &  \multicolumn{1}{c}{$f$ [\cday]} & \multicolumn{1}{c}{$a$ [mmag/\kms]} & \multicolumn{1}{c}{$\phi$ [0..1]} \\ \hline

\wire         & $0.251732(14) $ &  $12.26(18) $ &  $0.2195(34)$    \\  
\smei         & $0.2517553(34)$ &  $12.63(13)$  &  $0.2158(22)$    \\
\spec         & $0.2517506(43)$ &  $ 0.998(14)$ &  $0.1550(34)$    \\ \hline 
\dinshaw      & $0.25174(14)$   &  $ 0.808(58)$ &                  \\  
\hatzes       & $0.251722(22)$  &  $ 0.781(22)$ &                  \\  
\hline 


\end{tabular}
\end{table}
\section{Amplitude change of the main mode}
\label{sec:fou}

To analyse the time series in the frequency domain, 
we calculated the Fourier amplitude spectra for each dataset.
In the \topp\ \panels\ in Fig.~\ref{fig:amp} we show the
amplitude spectra for the photometry from the \wire\ (red) 
and \smei\ (green) and the radial velocities from \spec\ (black). 
The \inset\ shows the prominent peak around 0.25\,\cday, corresponding to the known 4-day period.
Due to the long gaps between the three \wire\ datasets the spectral
window shows a more complex pattern than the nearly continuous datasets
from \smei\ and \spec.

We fitted a single sinusoid to the dominant peak
and the results for each dataset are given in Table~\ref{tab:per}.  
We list the phase of the \wire, \smei\ and \spec\ datasets,
relative to the zero point $t_0 = 2,453,000$.
It is seen that the frequencies are in very good agreement.
Although the point-to-point precision of the \wire\ data is superior,
the long span of the \smei\ and \spec\ datasets means 
the frequency is determined more accurately by a factor of 3--4.
We also note that the phases from the fit to 
the \smei\ and \wire\ photometry are in good agreement,
but the phase of the fit to the \spec\ data is quite different.
This corresponds to a shift in the times of maximum 
in the flux and radial velocity:
${\rm HJD}(F_{\rm max})={\rm HJD}(RV_{\rm max})+(1.744\pm0.016\,{\rm d})$.
In comparison, \cite{moska00} found the offset to be $1.682\pm0.017\,{\rm d}$
using combined photometry and radial velocity data from \cite{kamper98}.
The shifts appear to be marginally different ($2.7\,\sigma$), 
but we note that the combined \cite{kamper98} data is calibrated to Johnson $V$, 
while \smei\ has a filter response roughly centered at Johnson $R$ but being wider \citep{tarrant07}.


We have subtracted the main oscillation mode
and the residual amplitude spectra are shown in the \bott\ \panel\ in Fig.~\ref{fig:amp}.
All spectra show a higher amplitude towards lower frequencies.
This could be either due to instrumental drift or long-period variation intrinsic to \polaris,
which we discuss in Sect.~\ref{sec:rv}.
The \lee\ \inset\ shows that the peaks in the \smei\ data
do not coincide with peaks in the \spec\ data,
so we cannot claim they are due to coherent pulsations.
The \rii\ \inset\ shows the details around $0.25$\,\cday, where
two significant residual peaks are seen in the \smei\ and \spec\ datasets.
They are an indication that the frequency, amplitude or phase of
the oscillation has changed during the $\sim4$ years of observation.

%
   \begin{figure}
   \centering
   \includegraphics[width=8.3cm]{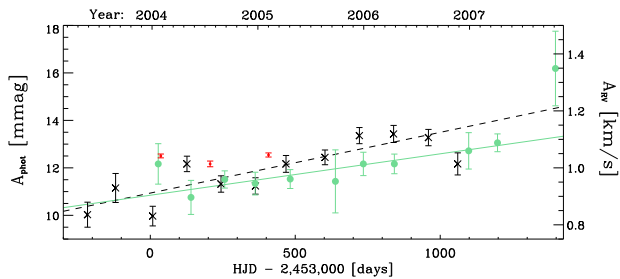}
      \caption{Increase in the amplitude of the 4\,d main mode measured over $4.5$ years 
               with \wire\ (\wiresymb), \smei\ (\smeisymb), and \spec\ (\specsymb).
         \label{fig:ampvar}}
   \end{figure}
%

To test the possible change in either frequency, amplitude or phase of the
main mode, we divided the datasets into subsets. 
Each of the \smei\ and \spec\ subsets contain 30 pulsation cycles, 
and the three \wire\ datasets were analysed independently.
Each subset is fitted by a single sinusoid and we analysed the output parameters.
We made the ana\-lysis for two different assumptions: 
\begin{itemize}
\item We assumed that the frequency, amplitude and phase change with time, so each is a free parameter.
\item We assumed that only the amplitude changes with time, so the frequency and phase were held fixed.
\end{itemize}

The first assumption allowed us to test the stability of the mode.
We found that within the uncertainties both the frequency and phase 
do not change over the time span of the observations. 
However, the amplitude increases monotonically, and
this result is confirmed by the analysis under the second assumption.
A comparison of the measured amplitudes under the two assumptions
give nearly identical amplitudes for all subsets.

In Fig.~\ref{fig:ampvar} we show the result obtained 
under the second assumption, \ie\ for fixed frequency and phase. 
The \smeisymb\ symbols are the \smei\ data, 
the \specsymb\ symbols are the \spec\ data, 
and \wiresymb\ are for the three \wire\ runs. 
The dashed line is a weighted linear fit to the \smei\ data yielding
the amplitude change in mmag: $A_{\rm phot}(t) = (10.94\pm0.08) + (2.56\pm0.13)\cdot10^{-3} (t-t_0) $, 
where $t$ is the HJD with zero point $t_0=2,453,000$. 
The weighted fit to the \spec\ data (solid line) gives:
$A_{RV}(t) = (0.90\pm0.01) + (1.45\pm0.15)\cdot10^{-4} (t-t_0)$\,\kms.
From this analysis we find that from 2003--7 the amplitude measured
in flux and radial velocity has increased by $34\pm2$\% and $24\pm3$\%, respectively.
The single mode in \polaris\ has very low amplitude and from linear theory it is
expected that the rate of increase is the same in photometry and radial velocity. 
The measured increase over four years is only marginally different ($8.1\pm3.1$\% or $2.6\,\sigma$), 
and continued monitoring is required to confirm this tentative result.




%

\begin{figure}
   \centering
   \includegraphics[width=8.9cm]{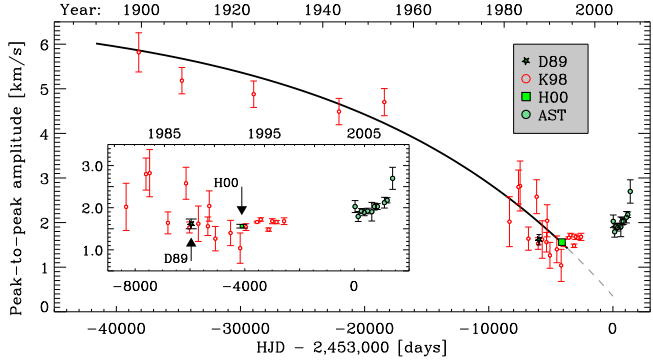}
      \caption{Peak-to-peak radial velocity amplitude for the main 4\,d mode.
               The solid line is the fit from \cite{kamper98}.
               The \inset\ shows the most recent data in detail. 
               Our measurements of the amplitude
               based on the \cite{dinshaw89} and \cite{hatzes00} are marked by arrows.
         \label{fig:kamper}}
   \end{figure}

In Fig.~\ref{fig:kamper} the measured peak-to-peak amplitudes 
in radial velocity from \spec\ are compared with results from the literature.
\cite{kamper98} found a monotonic decrease in the amplitude 
over the past 100 years\footnote{\cite{kamper98} converted 
their radial velocity amplitudes 
to photometric amplitudes using an empirical factor 50\,\kms\,mag$^{-1}$.
The ratio of the amplitudes measured in our radial velocity and the \smei\ photometric data 
is $A_{\rm RV}/A_{\rm phot}=79.0\pm1.4$\,\kms. 
The reason for the value being higher is that the filter response 
of \smei\ is roughly Johnson $R$ while \cite{kamper98} used Johnson $V$.}
To avoid confusion about the conversion factor, we compare our results directly in radial velocity,
and we have reproduced their fit as the solid black line.
The dashed line is an extrapolation, and as noted by \cite{kamper98}, 
zero amplitude would be predicted in the year 2007.
However, from four years of monitoring (1993.15--1996.96) in radial velocity, 
\cite{kamper98} found that the peak-to-peak amplitude was nearly constant.
This is in agreement with the amplitudes we find from the analysis (see Sect.~\ref{sec:specres}) 
of the datasets by \cite{dinshaw89} and \cite{hatzes00}. 
From the most recent high-precision data, it is evident that the amplitude in \polaris\
was constant in the period 1987--1997, while the increase we report here
marks a new era in the evolution of \polaris.

%
%

\subsection{Uncertainties in frequency, amplitude and phase\label{sec:uncert}}

The uncertainties in the frequency, amplitude and phase in Table~\ref{tab:per} 
are determined from realistic simulations of each dataset.
The times of observation are taken from the observations and 
the simulations take into account the increase in noise towards low frequencies. 
We used the approach described by \cite{bruntt07m67} (their Appendix~B)
to make 100 simulations of each dataset, 
and the uncertainties are the \rms\ value on the extracted frequency, amplitude and phase.

We made simulations of the \wire, \smei\ and \spec\ datasets 
and also the radial velocity datasets from \cite{dinshaw89} and \cite{hatzes00},
which we use for comparison in Sect.~\ref{sec:specres}. 
In all cases the uncertainties are somewhat higher than theoretical 
estimates that assume the noise to be white \citep{mont99}. 
This is especially the case for the \wire\ photometry due 
to low-amplitude variation on timescales comparable to the 4\,d period,
as will be discussed in Sect.~\ref{sec:wireres}.

%
   \begin{figure}
   \centering
   \includegraphics[width=8.8cm]{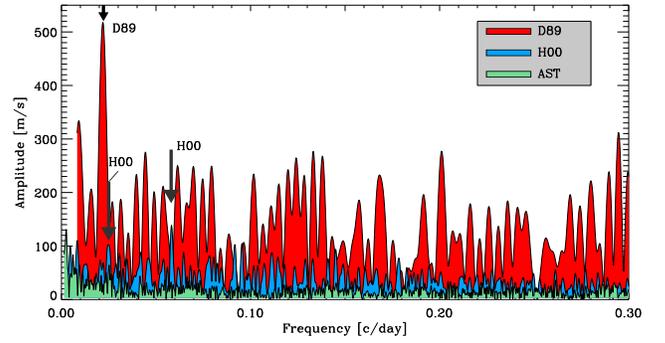}
      \caption{Residual amplitude spectra of the three radial velocity datasets 
from \spec\ (black), Hatzes \& Cochran (2000; blue), and Dinshaw et al. (1989; red). 
The low frequency peaks claimed in the previous studies are marked by arrows.
         \label{fig:specres}}
   \end{figure}
%
%

\section{Evidence for additional intrinsic variation}
\label{sec:rv}

In the following we will look
at the evidence for variation in \polaris\ beyond the 4\,d main mode.
The analysis is based on the \spec\ and \wire\ datasets and
data from two previous radial velocity campaigns. 
The \spec\ dataset is the most accurate at low frequencies, 
and can be used to look for long periods (Sect.~\ref{sec:specres}).
The \wire\ dataset consists of three week-long runs with very high precision
and can be used to study low-amplitude variation (Sect.~\ref{sec:wireres}).
The noise in the \smei\ dataset is distinctly non-white, 
as mentioned briefly in Sect.~\ref{sec:obs}. 
Therefore, we did not use this dataset in the following investigations,
except for the subtraction of the 4\,d mode, where \smei\ provides the 
most accurate frequency and phase.

\subsection{Long-period variation\label{sec:specres}}

In addition to the 4\,d mode,
a few studies have claimed variation at long periods, \eg\
$P=9.75$\,d \citep{kamper84}, $45.3\pm0.2$\,d \cite{dinshaw89}, 
and $40.2\pm0.7$\,d and $17.2$\,d \citep{hatzes00}. 
To confirm these claims we reanalysed 
the original datasets of \cite{dinshaw89} and \cite{hatzes00}.
The basic properties of these datasets are listed in Table~\ref{tab:data}.

In Fig.~\ref{fig:specres} we compare the amplitude spectra of the spectroscopic
data from \spec, \cite{dinshaw89} and \cite{hatzes00},
after subtracting the 4\,d period and the long-period trend due to the binary orbit. 
The \spec\ spectrum was calculated taking 
the amplitude increase into account (\cf\ Sect.~\ref{sec:fou}).
As a result, the residual double peak seen at $\sim 0.252$\,\cday\ is no longer visible
(compare Fig.~\ref{fig:specres} with the \bott\ \panel\ in Fig.~\ref{fig:amp}).
In the \spec\ amplitude spectrum we see no significant peaks 
from $0.02$--$0.3$\,\cday\ ($P=3$--$50$\,days).
We set an upper limit on the amplitude at 100\,\msec,
which is four times the average level in the amplitude spectrum in this frequency interval.

The dataset by \cite{hatzes00} consists
of 42 data points distributed unevenly over 640 days,
providing a very complicated spectral window.
The dataset comprises very precise radial velocities collected with an iodine cell 
as a reference (\rms\ residuals are 70\,\msec).
\cite{hatzes00} detected two peaks with almost equal amplitude at 
$f_{\rm H00}^{\rm A}=0.0249$ and $f_{\rm H00}^{\rm B}=0.0581$\,\cday.
In the amplitude spectrum in Fig.~\ref{fig:specres} 
we find only one of these peaks to be significant
($f_{\rm H00}^{\rm B}=0.05834\pm0.00014$\,\cday\ with amplitude $A=163\pm25$\,\msec).
We made 100 simulations of the data including this frequency,
white noise, and a $1/f$ noise component consistent with the observations.
The inserted $f_{\rm H00}^{\rm B}$ frequency was only recovered in 49 of the 100 simulations. 
Interestingly, in the other simulations the highest peaks 
were clustered close to $f_{\rm H00}^{\rm A}$ ($0.025\pm0.007$\,\cday\ with amplitude $A=167\pm22$\,\msec),
indicating that the two peaks detected by \cite{hatzes00} are due to the complicated spectral window.

The dataset by \cite{dinshaw89} comprises 175 velocities distributed over 241 days.
The precision is 440\,\msec, 
as estimated from the \rms\ of the residuals after subtracting the 4\,d mode.
In addition to this mode we detect a peak at $1.02199\pm0.00024$\,\cday\ with strong aliases at $\pm1$\,\cday. 
\cite{dinshaw89} found one of the alias peaks ($f_{\rm D89}=0.0221\pm0.0001$\,\cday) 
to be the highest and suggested that it was intrinsic to \polaris.
We measure the amplitude to be $A=0.542\pm0.067$\,\kms, but
such a strong signal is not seen in the more recent dataset 
by \cite{hatzes00} or in any of our datasets (see Figs.~\ref{fig:amp} and \ref{fig:specres}). 
For these reasons we believe that this peak is a 1\,\cday\ artifact, 
likely caused by instrumental drift in combination with the spectral window.


%
   \begin{figure}
   \centering
   \includegraphics[width=8.8cm]{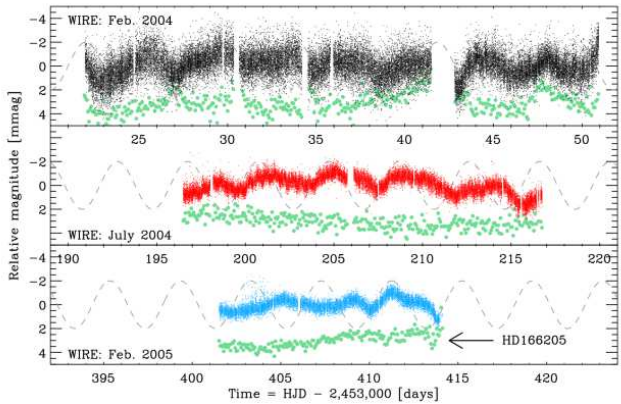}
      \caption{Residuals in the three \wire\ light curves of \polaris\
               and the comparison star HD~166205 (large $\bullet$ symbols offset by 3\,mmag).
               The dashed curve is a sinusoid with the 4\,d period and an amplitude 
               set arbitrarily to 2\,mmag. 
         \label{fig:wireres}}
   \end{figure}
%
%

%
\subsection{Low-amplitude variation \label{sec:wireres}}
%

The residuals in the \wire\ light curves,
after having subtracted the 4\,d period, are shown in Fig.~\ref{fig:wireres}.
The drifts seen in the \wire\ light curves correspond to periods from 2--6\,d,
\ie\ similar to the 4\,d main mode. Before interpreting this variation, 
it is important to investigate if these variations 
are due to improper subtraction of the main period or due to instrumental drift.

To test the first caveat, we compared the residuals 
when subtracting the fit to the \smei\ data and 
when subtracting fits to the individual \wire\ datasets.
We find that the residuals are quite similar, and the conclusions reached here
are not affected by the adopted approach.
In the following, we have used the most accurate 
value for the frequency and phase, which is from the fit of the main mode to the \smei\ data,
while we used the amplitudes fitted to the individual \wire\ datasets.

The second caveat is instrumental drift and we tested this by using a comparison star.
During the observations with the \wire\ star tracker 
four other stars were monitored on the same CCD.
However, not all are suited as comparison stars:
HD~5848 is a bright ($V=4.3$) K~giant star clearly showing solar-like oscillations \citep{stello08}.
HD~51802 and HD~174878 are faint ($V=5.1$ and $6.6$) M~giants showing variation with long periods. 
The only suitable comparison star is the A1\,V star HD~166205 ($\delta$~UMi; $V=4.4$). 
In Fig.~\ref{fig:wireres} we also show the 
light curve of this star with large $\bullet$ symbols. 
The star is two magnitudes fainter than \polaris, 
so we binned the data collected within each orbit ($P_{\rm orbit}\simeq93$\,min)
to be able to see any low-amplitude variation.

   \begin{figure}
   \centering
   \includegraphics[width=8.8cm]{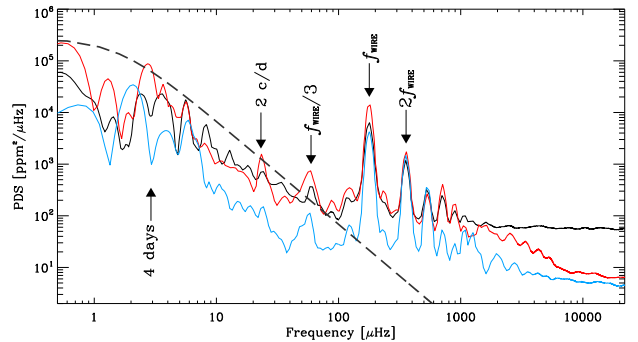}
      \caption{Power density spectrum (PDS) 
of the residual light curves from \wire\ shown in Fig.~\ref{fig:wireres}. 
The dashed line is a scaling of the granulation measured in the Sun.
The arrows mark the 4\,d main period and 
frequencies that are due to the \wire\ orbit.
         \label{fig:density}}
   \end{figure}

The clear variation in the residuals of the \wire\ photometry is not 
seen in the hot comparison star HD~166205. The fact that it is seen in three
light curves obtained with \wire\ at three different epochs spanning one year gives
us some confidence that the signal is intrinsic to \polaris.
We will discuss two possible explanations for the observed variation here, 
namely granulation and star spots.

To investigate in detail how the variation in the \wire\ light curves
depends on frequency, we show their power density spectra (PDS) in Fig.~\ref{fig:density}
in a logarithmic plot.
The virtue of the PDS is that one can directly compare the properties of 
datasets, which have different temporal coverage and noise characteristics.
Each PDS shows a clear increase from the white noise level around $10$\,mHz 
towards low frequencies.
The arrows mark the harmonics of the orbital frequency of \wire\ ($f_{\rm WIRE}=178$\,\mhz),
one third of the \wire\ orbital frequency, and 2 c/day. These frequencies
are observed in almost all \wire\ datasets, and are due to a combination
of a low duty cycle (typically 20--40\%) and scattered light from earth shine.
Also marked is the location of the 4\,d mode ($f=2.91$\,\mhz), which has been subtracted.

To see if granulation could explain the increase in the PDS towards low
frequencies, we have used a scaling of the granulation observed in the Sun.
This is based on \virgo\ satellite observations of the Sun-as-a-star, 
and the scaling is done both in terms of amplitude and timescale,
following Kjeldsen \& Bedding (in preparation; see also \citealt{stello07}). 
This prediction is shown as the dashed line in Fig.~\ref{fig:density}. 
This is a scaling over three orders of magnitude in luminosity from 
the Sun to the super giant \polaris, so it is intriguing 
that the prediction of the granulation signal
agrees with the observations within a factor of $3.0\pm0.5$ in amplitude.

Both \cite{dinshaw89} and \cite{hatzes00} discussed evidence for
star spots on \polaris. We argue that this is not the cause of the observed 
variation in the \wire\ data, since the time-scale is only a few days. 
That would imply very rapid rotation, which is unlikely for such an evolved star.

\section{Discussion and conclusion}

From $4.5$ years of intensive monitoring of \polaris\ we find that
the amplitude of the 4-day main mode has increased steadily by about 30\% in 
both radial velocity and flux amplitude.
The rate of increase in the amplitude from 2003--2007 is slightly steeper than the decrease from the
fit by \cite{kamper98} for the period 1980--1994.
Other sources have also found  a recent increase in the amplitude of \polaris\ 
(\citealt{engle04}; Turner 2007, private communication).  

The result that the amplitude of \polaris\ is now increasing has implications
for earlier explanations of the change in amplitude as a cessation of pulsation 
due to the Cepheid's evolution towards the edge of the instability strip.
At the very least, whatever the process is, 
it is not a simple monotonic progression through the HR diagram,
unless this stage of evolution is more complex than previously thought.
The recovery of the amplitude suggests that the phenomenon is cyclic.
As such, it is likely to be associated in some way with pulsation
rather than with evolution.

A possible explanation for the increase in amplitude could be the beating
of two closely spaced modes \citep{breger06}.  To exhibit a beat period as long
as the amplitude variation of \polaris\ (likely a few 100 years), the modes
would have to be very closely spaced indeed. Among classical Cepheids,
amplitude variation is extremely uncommon.  The only case where it is 
firmly established is in  V\,473~Lyrae \citep{burki86}, where the variation
occurs over only a few years.  In RR~Lyrae stars, 
the Blazhko effect is well known, and thought to be the result of mode beating,
although it is  not completely understood.  Since the amplitude changes in \polaris\  
are well established, they require further observations and 
consideration theoretically to unravel the cause. 

There are other interesting aspects to the main oscillation mode.
The characterization of the period change is puzzling
since analysis of $O-C$ diagrams
spanning more than a century reveal that the period change is not linear. 
There may have been a ``glitch" in both the period change and the
amplitude in the mid-1960's \citep{turner05}, rather than smooth changes. 
We measured 239 epochs of maximum light but the time span of the observations
of $4.5$ years is too short to investigate the period change.
This is another aspect of the pulsation that demands further observation. 

A few radial velocity campaigns have claimed the presence of 
additional long-period variation in \polaris. We have analysed the original
datasets by \cite{dinshaw89} and \cite{hatzes00} 
and we conclude that these long periods
are likely spurious detections caused by instrumental drifts \citep{dinshaw89}
or insufficient data leading to a complicated spectral window \citep{hatzes00}.
From our 3.8\,yr of monitoring with \spec, we set an upper limit 
on the variation in radial velocity at 100\,\msec\ for periods in the range 3--50\,d 
(except for the 4\,d main mode).

In the \wire\ data we find evidence of low-amplitude variation (peak-to-peak $2$\,mmag)
at time scales of 2--6 days, which are likely intrinsic to \polaris.
We applied a simple scaling of observed solar values for the characteristic 
timescale and amplitude of the granulation. 
Although this is a scaling over three orders of magnitude in luminosity, 
the prediction agrees with the observed variation in \polaris\ within a factor $3.0\pm0.5$.
Thus, we conclude that
the variation in the \wire\ data could be due to granulation.

\acknowledgments

HB and DS are supported by the Australian Research Council.
NRE and AJP acknowledge support from NASA contracts NAS8-03060 and
NNG~05GA41G, respectively. Operation of the Tennessee State University
Automatic Spectroscopic Telescope was supported by grants from NASA
(NCC5-511) and NSF (HRD 9706268). \smei\ was designed and constructed by
a team of scientists and engineers from the at University of California
at San Diego, Boston College, Boston University, and the University of
Birmingham. 
We thank A.~Buffington, C.~J.~Eyles and S.~J.~Tappin for advice 
on the \smei\ data reduction.
The following internet-based resources were used for this paper: 
the \nasa\ Astrophysics Data System
and the ar$\chi$iv scientific paper preprint service operated by Cornell University.

\bibliography{ms}

\end{document}